# Reconfigurable interactions and three-dimensional patterning of colloidal particles and defects in lamellar soft media

Rahul P. Trivedi[a,b,c], Ivan I. Klevets[a,d], Bohdan Senyuk[a], Taewoo Lee[a], and Ivan I. Smalyukh[a,b,c,e,*]

[a]*Department of Physics, University of Colorado, Boulder, CO 80309*
[b]*Liquid Crystal Materials Research Center, University of Colorado, Boulder, CO 80309*
[c]*Department of Electrical, Computer and Energy Engineering, University of Colorado, Boulder, CO 80309*
[d]*Institute for Condensed Matter Physics of the National Academy of Sciences of Ukraine, 1 Svientsitskii Street, 79011 Lviv, Ukraine*
[e]*Renewable and Sustainable Energy Institute, National Renewable Energy Laboratory and University of Colorado, Boulder, CO 80309*

*\*E-mail: ivan.smalyukh@colorado.edu*

**Abstract**

Colloidal systems find important applications ranging from fabrication of photonic crystals to direct probing of phenomena typically encountered in atomic crystals and glasses. New applications - such as nanoantennas, plasmonic sensors, and nanocircuits - pose a challenge of achieving sparse colloidal assemblies with tunable interparticle separations that can be controlled at will. We demonstrate reconfigurable multiscale interactions and assembly of colloids mediated by defects in cholesteric liquid crystals that are probed by means of laser manipulation and three-dimensional imaging. We find that colloids attract via distance-independent elastic interactions when pinned to the ends of cholesteric oily streaks, line defects at which one or more layers are interrupted. However, dislocations and oily streaks can also be optically manipulated to induce kinks, allowing one to lock them into the desired configurations that are stabilized by elastic energy barriers for structural transformation of the particle-connecting defects. Under the influence of elastic energy landscape due to these defects, sublamellar-sized colloids self-assemble into structures mimicking the cores of dislocations and oily streaks. Interactions between these defect embedded colloids can be varied from attractive to repulsive by optically introducing dislocation kinks. The reconfigurable nature of defect-particle interactions allows for patterning of defects by manipulation of colloids and, in turn, patterning of particles by these defects, thus achieving desired colloidal configurations on scales ranging from the size of defect core to the sample size. This defect-colloidal sculpturing may be extended to other lamellar media, providing the means for optically guided self-assembly of mesoscopic composites with predesigned properties.

**Introduction**

Topological defects play an important role in determining material properties, mediating phase transitions, and causing numerous irreversible phenomena (1–6). For example, dislocations - line defects in periodically ordered materials - impart malleability to solids such as metals and thus are of great practical utility. Liquid crystals (LCs), soft materials with long-range orientational but only partial or no positional order (1, 7), are often employed as model systems to study defect-related phenomena in fields ranging from early universe cosmology (8) to condensed matter (9, 10). For example, lamellar media are often studied using the model system of a cholesteric LC in which the constituent chiral molecules locally align along a common direction called "director" $\mathbf{n}(\mathbf{r})$ that twists around a helical axis to form a twist-periodic structure (7, 11–13). When spatial distortions of the equilibrium twisted structure are on a scale much larger than the helical periodicity, this one-dimensionally periodic system exhibits elastic properties similar to those of layered systems such as smectic LCs (7, 11).Molecular, atomic, or spin configurations periodic in one direction have many similar properties and are encountered in a multitude of other condensed matter systems, such as helimagnets (13), various smectic phases of LCs (1, 10), water-surfactant mixtures (14, 15), block-copolymers (16), colloidal assemblies (17), lipids (18), DNA condensates (19), etc. These media are of great interest from the standpoint of controlling their bulk elastic and optical properties by stabilizing and decorating networks of defects with particles (20–22). A number of studies of colloids in LCs have revealed how their properties are enriched by presence of defects (20–33). The fascinating richness of interactions involving colloids and defects in lamellar systems (14, 20, 21, 27, 28, 32, 33) and their potential practical applications require that these effects be explored on the level of individual particles and defects.

In this work, we probe defect-particle interactions and show that colloidal stabilization of defect structures can be extended to individual defect-particle level. Because the nature and strength of these interactions are defect-dependent, we use three-dimensional (3D) optical imaging based on fluorescence confocal polarizing microscopy (FCPM) (34) and two-photon excitation fluorescence polarizing microscopy (2PEF-PM) (35) (Fig. S1) to visualize the internal defect core structures and layer deformations. We optically generate various defects and alter their internal structure via noncontact manipulation with laser tweezers and then quantitatively probe the ensuing particle-defect and particle-particle interactions. We demonstrate the feasibility of forming predesigned stable 3D configurations of colloidal particles arranged along defects within cholesteric layers as well as along directions perpendicular to the layers. This may enable the physical "synthesis" of novel composites having optically controlled properties, with patterned defect lines playing the role of artificial bonds between colloids and providing a framework for testing various elastic theories. We also demonstrate the ability to pattern sublamellar-sized particles within the defect cores, which is of great interest for nanophotonics applications.

**Colloidal Interactions Mediated by Topologically Unstable Defect Clusters**

Cholesteric LCs host a great variety of linear defects that are also encountered in other lamellar systems (1, 2). The most abundant of these defects are dislocations and oily streaks that terminate or interrupt one or more layers (10). Elementary edge dislocations have Burgers vector $|\mathbf{b}| = b = p/2$ (where $p$ is the cholesteric pitch, the distance along the helical axis over which $\mathbf{n}(\mathbf{r})$ twists by $2\pi$) (Table S1). Their cores comprise a dipole of $\lambda$- and $\tau$-disclinations of opposite half-integer strength.[*] An elementary oily streak (also called "Lehmann cluster") has the net Burgers vector equal to zero and is composed of a quadruple of nonsingular $\lambda$-disclinations, viz., two $\lambda^{+1/2}$, and two $\lambda^{-1/2}$ disclinations, Fig. 1. Unlike dislocations, Lehmann clusters can terminate on colloidal particles because they have the same number of cholesteric layers on opposite sides and structure compatible with tangential surface anchoring conditions for $\mathbf{n}(\mathbf{r})$ on the colloidal spheres such as the used melamine resin polymer microparticles (see Materials and Methods for details). The structures of defect-connected colloids like one shown in Fig. 1 A, B, and D can occur spontaneously during defect coarsening upon quenching the LC cell from isotropic phase. These defect-particle configurations can also be generated using laser tweezers (Movie S1). These defects mediate distance-independent attractive interaction between the particles, with the interaction force equal to the defect line tension (free energy per unit length) (Fig. 1). We measure this force via optical manipulation of a particle attached to the tip of the line defect and tracking its motion, upon being released from the trap, and being dragged at a constant velocity $v$ (Fig. 1E and Movie S2). Owing to relatively small $v$ and high effective viscosity $\eta$ of the LC (i.e., the Reynolds number is small), the inertial forces acting on the particle are negligible. The defect-mediated pair-interaction force $F_{\text{int}}$ is balanced by the viscous drag that can be estimated using Stokes' law, yielding $F_{\text{int}} = F_{\text{viscous}} = 6\pi\eta R v$, where $R$ is the particle radius. We use a value of viscosity coefficient that is the average of the measured values for directions parallel and perpendicular to the uniform $\mathbf{n}(\mathbf{r})$, obtaining $\eta = 56.3$ mPa s for 5CB-based cholesteric (36). Interactions due to defect lines of zero Burgers vector may involve multiple mobile or stationary particles interconnected by defects. For example, Fig. 1B shows a node of three Lehmann clusters, sustained with their ends pinned to particles. The particle released from a laser trap is dragged by the receding defect with the ensuing force equal to its line tension (Fig. 1E and Movie S3).

---

[*] In the $\lambda$ disclinations, the material $\mathbf{n}$-director field is non-singular because of $\mathbf{n}$ being parallel to the defect line in its core, so that the singularity is observed only in the immaterial director fields $\chi$ (along the helical axis) and $\tau$ orthogonal to both $\mathbf{n}$ and $\chi$. In the $\tau$-disclinations, the $\tau$ field is non-singular but $\mathbf{n}$ and $\chi$ director fields are singular. In $\chi$-disclinations, the singularities are found in both $\mathbf{n}$ and $\tau$ director fields but not in the $\chi$ field (10).

The strength of the distance-independent defect-mediated interactions depends on the core structure of oily streaks (Figs. S2–S6). For example, the fragments of optically manipulated Lehmann cluster shown in Fig. 2A turn by ±90°, so that each part shifts vertically by a distance of ±$p$/4, to conform to the cholesteric helix. The Lehmann cluster stretched thus metamorphoses into a defect with a different core structure schematically shown in Fig. 2C, as reconstructed using FCPM vertical cross-sections (Fig. 2 D and E). Although the ensuing elastic force acting on this defect-dragged microsphere is also distance-independent, its value is about 1.3–1.4 times larger than for the original Lehmann cluster (Movie S4). Similar study of other oily streaks (Fig. S6) reveals that the strength of defect-mediated colloidal interactions increases with their width. Computer simulations based on minimization of the Frank elastic free energy yield minimum-energy internal structure, line tension, and cross-sectional images of oily streaks consistent with the experiments (Fig. S4 and Table S2). For example, line tension values for the Lehmann cluster in 5CB-based cholesteric obtained by experimental methods range from 17.0 pN to 19.6 pN and the numerical estimate is 20.5 pN.

Lehmann clusters can also "glide" across the layers via "kinks" mediated by winding of the constituent $\lambda^{+1/2}\lambda^{-1/2}$ disclinations (Fig. 3). Fig. 3A shows a Lehmann cluster terminating on a particle and traversing across several layers via kinks such that each winding shifts the cluster across the layers by distance $p$. In-plane images shown in Fig. 3 B–D and cross-sectional images in Fig. 3 E–H, obtained in 2PEF-PM imaging mode, show the vertical displacement of the dislocation due to the kink with winding of $\lambda$-disclinations. The schematic diagram in Fig. 3I shows reconstructed $\mathbf{n}(\mathbf{r})$ and traces the disclination lines $\lambda^{+1/2}$ and $\lambda^{-1/2}$ that preserve their nonsingular nature by conforming to the cholesteric helix while winding. Both the kink and the colloidal particle weakly pinned to it (with the binding energy approximately 20–100$k_B T$) can be optically translated along the length of the defect (Fig. 3J). Colloidal particles placed at the location of kinks and anti-kinks allow one to affix their positions and prevent their mutual annihilation, thus achieving controlled patterning of colloids and defects in 3D. This shows that the colloidal stabilization of defect structures in cholesteric LCs, originally demonstrated by Zapototsky et al. (20), can be extended down to the level of individual defects and particles. Our experiments indicate that morphing of defect structures between different states typically involves consecutive transformations that preserve the nonsingular nature of defect cores and/or involve low-energy structures that are not separated by elastic energy barriers. Colloids placed at the nodes and kinks hinder these defect transformations by imposing strong energy barriers $\gg k_B T$ and entrap the system in long-lived metastable states.

**Defect-Arbitrated Colloidal Patterning**

Elasticity-mediated interactions between defects and particles can be exploited for their 3D patterning. While large particles typically localize at defect nodes (20, 27), colloids of size smaller than the layer thickness exhibit a wide variety of anisotropic interactions with defects and also with each other when confined in defect cores. To explore these interactions, we use melamine resin particles with $R = 1$ µm and tangential boundary conditions for $\mathbf{n(r)}$, dispersed in a cholesteric LC with $p = 8$ µm. These particles experience highly anisotropic interactions that drive them into well-defined positions within defect cores (Fig. 4) and yield binding potential of the order of $1,000 k_B T$. Depending on its initial location, each colloid can stably colocalize with any of the four $\lambda$-disclinations forming a Lehmann cluster (Fig. 4 A–D). For example, the attraction of a particle toward the $\lambda^{+1/2}$ disclination is within a range of approximately $4R$ (Fig. 4E), with the distance dependence of the interaction energy shown in Fig. 4F.

Defects and the corresponding director distortions create an elastic potential landscape for the particles to interact while being localized within their cores. The interactions of particles centered on the same $\lambda$-disclinations of the core yield binding energies approximately $2,000 k_B T$. The particles centered on different disclinations of dislocation or oily streak cores interact to localize roughly in the same plane orthogonal to the defect length (Fig. 4 G and H and Fig. 5 I–K). This can be exploited for self-assembly of various colloidal structures replicating the defect core morphology - e.g., a diamond-like arrangement of four colloids centered on $\lambda$-disclination within the Lehmann cluster (Fig. 4G–I). Interestingly, to further minimize elastic energy in this defect-bound colloidal configuration, particles centered at the two $\lambda^{-1/2}$ disclinations get slightly displaced with respect to each other along the defect length (Fig. 4I), so that only three of the four particles are clearly visible in the cross-sectional images orthogonal to the defect line (Fig. 4 G and H). Two particles embedded in the same $\lambda$ disclination experience a relatively long-range (up to $\approx 12R$) attractive interaction along the defect line (Fig. 4 E and F).

Glide of dislocations across lamellae via kinks enables colloidal structuring in 3D (Fig. 5 A–D). Although particles centered on the same disclination of an elementary edge dislocation core attract each other along its length (Fig. 5 E and H and Movie S6), they repel from a kink and also from each other when separated by kinks. This behavior is different from the above-discussed attractive interaction between a similar colloid and a vertical winding kink in a Lehmann cluster with nonsingular defect structure (Fig. 3J). The kink in the edge dislocation has length and tilt with respect to lamellae that minimize the overall elastic energy cost of connecting edge dislocation fragments at different layers. When optically moved toward a kink, a particle alters the equilibrium structure, length, and tilt and thus is repelled from it to minimize elastic energy. The magnitude of interaction potential is of the order of $1,000k_BT$ at short distances. Therefore, these interactions can be employed to selectively arrange colloids at different height across lamellae while being centered at the $\tau^{\pm 1/2}$ or $\lambda^{\pm 1/2}$ lines. For example, particles embedded in the same elementary dislocation and separated by kinks form a sparsely spaced ladder-like structure shown in Fig. 5 B–D. The sequence of vertical cross-sections (Fig. 5D) demonstrates how the $b = p/2$ dislocation core alternates between that composed of $\tau^{-1/2}\lambda^{+1/2}$ and $\lambda^{-1/2}\tau^{+1/2}$ disclinations (Fig. 5 F and G) and how particles find well-defined equilibrium positions while interacting with each other and kinks within the core. Dislocations and oily streaks can be optically manipulated to induce kinks and thus generate a desired profile of elastic potential along their lengths, both within and across the layers. For example, the kink/anti-kink pair in the dislocation of Burgers vector $b = p/2$ shown in Fig. 5A is generated using optical tweezers acting on a pinned particle as shown in Fig. S7.[†] The axial optical trapping force needed to generate the kinks of $b = p/2$ dislocation via pulling or pushing colloidal particles across the layers is of the order of 10–15 pN, in agreement with the estimates of the Peierls–Nabarro friction due to the transformation between $\tau^{-1/2}\lambda^{+1/2}$ and $\lambda^{-1/2}\tau^{+1/2}$ disclination pairs in the core (6). The kink can be optically translated along the defect length (Fig. 5A), and their position can be affixed by use of particles larger than the lamellar size, as discussed above. Thus, the linear defects can be directed along a desired 3Dpath within and across lamellae, giving rise to the desired landscape of interaction potential for patterning of sublamellar-sized particles.

**Discussion**

Most colloidal self-assembly approaches rely on the formation of close-packed crystal structures,

---

[†] Inducing glide of $b = p$ dislocations requires transformation of the $\lambda^{-1/2}\lambda^{+1/2}$ non-singular disclination pair into a pair of singular $\tau^{-1/2}\tau^{+1/2}$ disclinations. The core energy increase under the transformation is approximately $pK\ln(p/r_c)$, where $r_c$ is the size of the singular core. For ZLI-2806 with $K\approx 12$ pN, one finds that the required force would be >100 pN, larger than the maximum axial trapping force value that can be exerted by the used optical traps. This explains why only the $b = p/2$ but not the $b = p$ dislocations can be optically shifted across layers via kinks.

taking advantage of typically short-range screened electrostatic and other interactions (1). We have demonstrated sparsely spaced controlled arrangements of colloids in cholesteric lamellae that utilize reconfigurable elastic interactions enriched by defects and stabilized by strong energy barriers for transformation and annihilation of defects caused by presence of particles. The nature of interactions and the intradefect assemblies are scale-invariant and hence can be controlled by changing the cholesteric pitch from about 50 nm to hundreds of microns. Because the defect structures and elasticity in other lamellar systems are similar to the ones studied here, our approach of using defects for sparse arrangement of colloids can be extended to other lamellar hosts such as diblock copolymers, DNA condensates, and various smectic LCs, as long as the energetic barriers due to the transformation of particle-stabilized defect structures can be kept $\gg k_\mathrm{B}T$. Importantly, because the observed behavior of colloids in this system only depends on the boundary conditions for $\mathbf{n}(\mathbf{r})$ at their surfaces, one can use particles of different material compositions, including metal plasmonic nanoparticles. Stable self-assembled arrangements of elastically bound two, four, or more particles centered in disclinations of the cores of dislocation and oily streaks are thus of great interest for applications such as nanoantennas and optical field concentrators, especially because the interparticle spacing can be varied from nanometers to microns. In addition to laser tweezers, the control of the 3D defect–particle architectures can be based on the application of electric fields, by using electrophoretic/dielectrophoretic forces applied to particles or through the modification of defect core structures due to direct coupling of $\mathbf{n}(\mathbf{r})$ with applied fields; using LCs with positive or negative dielectric anisotropy further enriches the control capabilities. The former approach offers tuning of line tension of linear defects like oily streaks (allowing one to even achieve negative tension values in materials with positive dielectric anisotropy), providing additional robust means for tuning interactions between particles at the ends of connecting defects or within the defect cores. Magnetic control can be efficient when using ferromagnetic or superparamagnetic colloids (37). One can expect that observed interactions would be highly dependent on the shape of used particles (38), possibly enabling selective shape-dependent localization of particles into different disclinations within the defect cores.

**Conclusions**

We have demonstrated 3D patterning of defect–colloidal networks in cholesteric lamellae by using optical generation and stabilization of desired defect structures by means of colloids. The stabilized defect networks, in turn, form a 3D elastic energy landscape capable of entrapping sublamellar-sized colloids. By introducing kinks in dislocations and oily streaks, interactions between these smaller colloids can be tuned from attractive to repulsive, offering powerful means to control sparsely spaced colloidal assemblies. Because both dielectric and metal particles can be used to form these defect-colloidal structures, potential applications include engineering of new structured materials, optical nanoantennas, and metamaterials. Acting as elastic "bonds" connecting particles in long-lived metastable states, these highly controlled defect networks also offer the means to design elastic properties of new LC–colloidal composites (20, 21, 39).

**Materials and Methods**

Cholesterics are prepared by mixing nematic LCs 4-Cyano-4′-pentylbiphenyl (5CB) or ZLI-2806 with a chiral dopant CB15 (all from EM Chemicals). The value of the pitch $p$ is set by choosing concentration of the chiral dopant ($C_{chiral}$) of known helical twisting power $h_{HTP}$ (Table S1), so that $p = 1/(h_{HTP} \times C_{chiral})$. For FCPM and 2PEF-PM, the LCs are doped with a small amount (0.05 wt%) of anisotropic dye n,n′-bis(2,5-di-tert-butylphenyl)-3,4,9,10-perylenedicarboximide (BTBP, Aldrich). We have used melamine resin particles of 2 and 4 μm in diameter. These particles set tangential boundary conditions for $\mathbf{n}(\mathbf{r})$ on their surface. We first obtain the particles in a powder form from an aqueous dispersion (from Aldrich) by evaporating water. The particles are then dispersed in the LC and the resultant mixture is sonicated to break particle aggregates. LC cells are made with two glass substrates of thickness 0.15 mm, as required by imaging and optical trapping with high-numerical aperture objectives. Strong planar surface anchoring boundary conditions are set by spin-coating and curing a thin layer of polyimide PI-2555 (HD Micro-Systems), and then unidirectionally rubbing it with a piece of velvet cloth to define the alignment direction. The thickness of LC cells is set to 30–60 μm by sandwiching the glass substrates with silica microspheres of corresponding size. The LC is infiltrated into the cells by use of capillary forces.

Optical manipulation and 3D imaging is performed with an integrated setup composed of holographic optical tweezers and multimodal imaging that includes FCPM and 2PEF-PM (see SI Text and Fig. S1). The use of two different 3D imaging techniques allows us to assure that various artifacts, such as the Mauguin following of light polarization in the twisted structures, do not affect the reconstruction of the $\mathbf{n}(\mathbf{r})$-structures. The defect tension is measured by several different approaches with the use of colloidal particles as "handles" and without them, all yielding consistent results described in details in the supporting material. Particle motion is tracked by videomicroscopy at 15 frames per second using a charge coupled devise camera.

**Acknowledgments**

This work was supported by the International Institute for Complex Adaptive Matter (R.P.T., I.I.K.) and National Science Foundation grants DMR-0645461 (I.I.S.), DMR-0820579 (R.P.T.), and DMR-0847782 (B.S., T.L., I.I.S.). We acknowledge discussions with Noel Clark and Randall Kamien.

# Figures

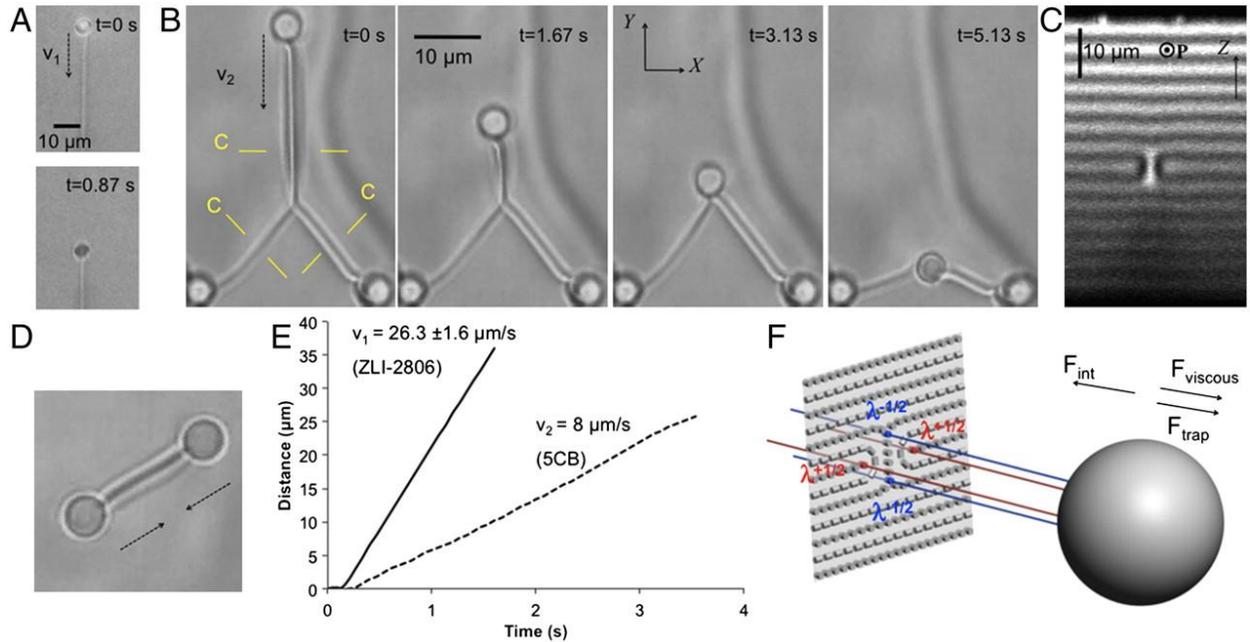

**Fig. 1.** Defect–particle interaction mediated by a Lehmann cluster. (A) A defect cluster in a cholesteric LC (ZLI-2806 and CB15) terminates at a particle and can be stretched by optically translating it but contracts once the colloid is released from the trap. (B) A Lehmann cluster in a cholesteric LC (5CB and CB15) is manipulated with the help of an optically trapped particle so as to form a node of clusters. Upon turning the trap off, the particle returns to its original position along the defect line between two stationary trapped colloids, as seen in the frames marked with elapsed time. (C) Vertical 2PEF-PM cross section of an elementary Lehmann cluster. (D) Lehmann cluster terminating on two particles. (E) Distance vs. time plot for the particle in A shown by solid line and for the particle in B shown by dashed line. (F) A schematic representation of an elementary Lehmann cluster terminating on a particle; the blue and red lines trace the $\lambda^{-1/2}$ and $\lambda^{+1/2}$ defects, respectively.

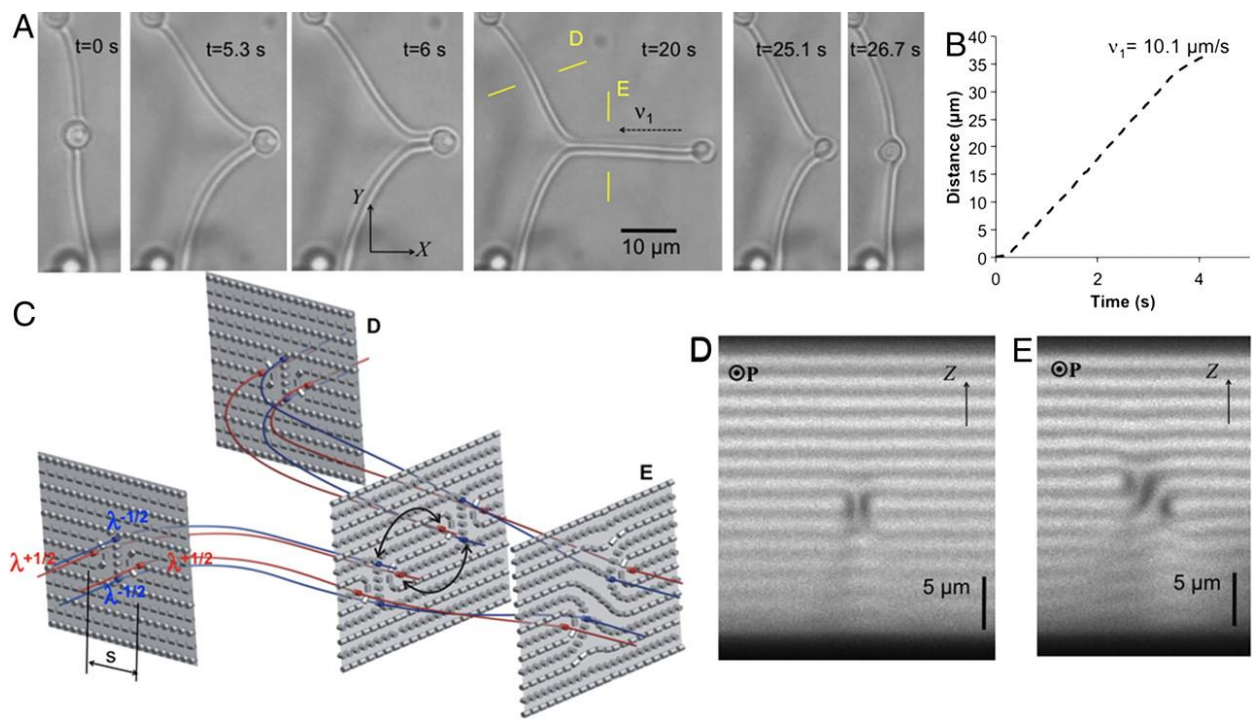

**Fig. 2.** Tensile defect–particle interactions mediated by various oily streaks. (A) An optically trapped particle, embedded in the core of an elementary Lehmann cluster, is used to transform the defect core structure. (B) Distance vs. time plot describing motion of the particle shown in A when released from the trap. (C) A schematic representation of the initial and the resultant defect core structures in their vertical cross-sections; the blue and red lines trace $\lambda^{-1/2}$ and $\lambda^{+1/2}$ disclinations in the defect cores, respectively. (D) FCPM vertical cross-sections of the Lehmann cluster and (E) the resultant oily streak having a modified core structure.

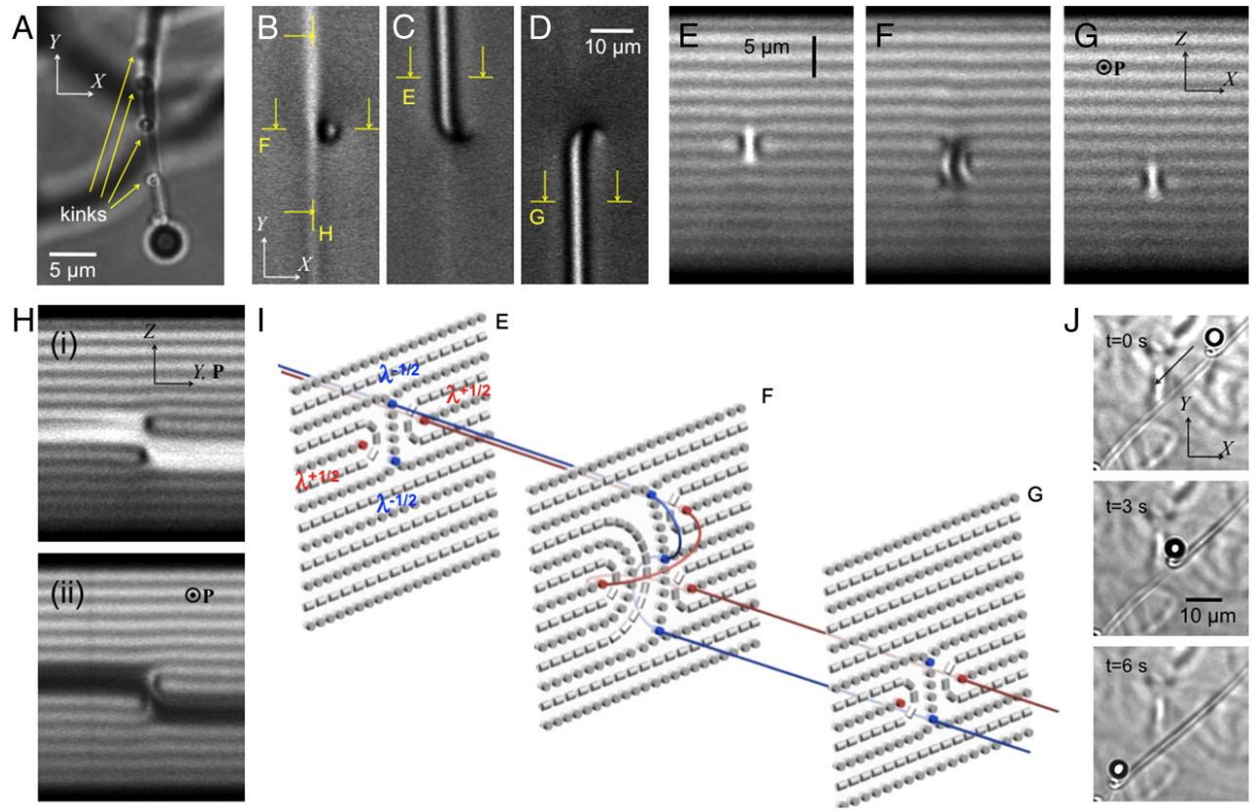

**Fig. 3.** Displacement of a Lehmann cluster across layers. (A) polarizing microscopy image showing how a Lehmann cluster terminating on an optically manipulated particle traverses layers via kinks in the form of windings. (B–D) In-plane 2PEF-PM images showing that the Lehmann cluster is at different depth across the layers on two sides of the kink. (E–G) Vertical 2PEF-PM cross-sections in the locations marked on the in-plane images (B–D). (H) Vertical 2PEF-PM cross-sections along the length of the defect for two orthogonal polarizations of the probing light. (I) A schematic of the reconstructed 3D director field and winding mechanism, with the vertical cross-section planes corresponding to cross-sectional images labeled alongside; the blue and red lines trace the $\lambda^{-1/2}$ and $\lambda^{+1/2}$ disclinations of the cluster, respectively. (J) The position of kinks along the cluster's length can be displaced or locked with the help of a particle.

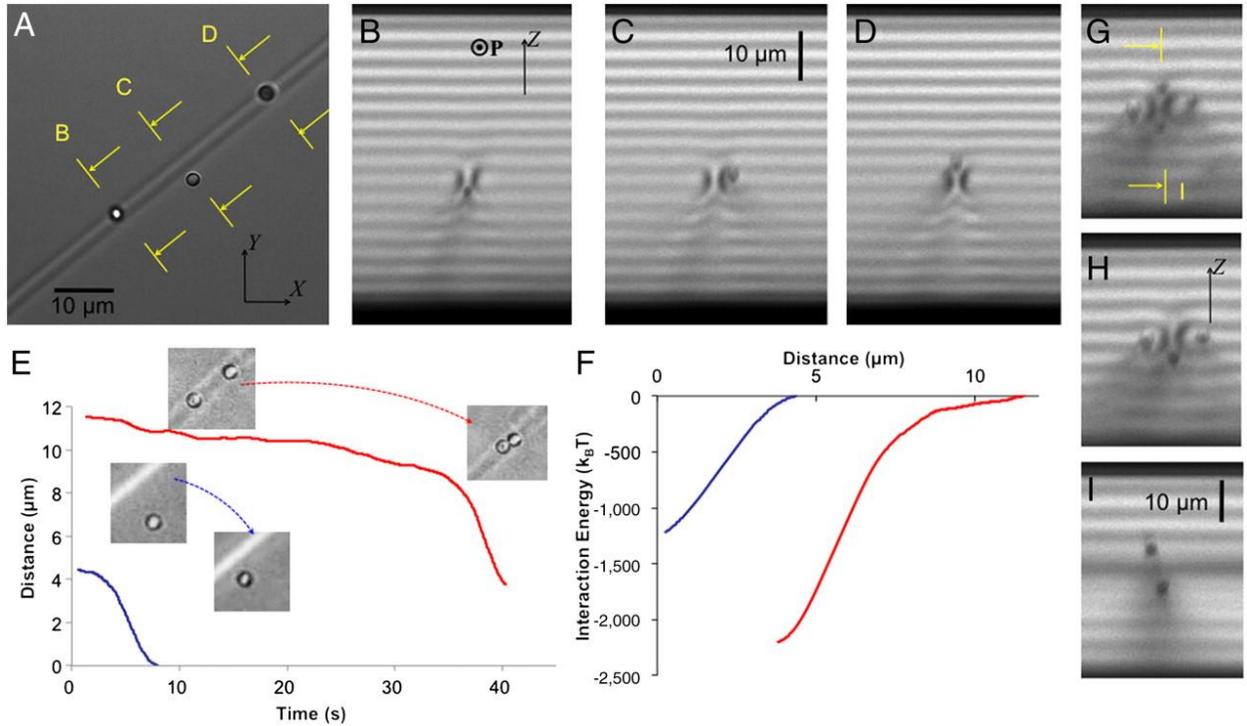

**Fig. 4.** Three-dimensional patterning of particles within a Lehmann cluster. (A) In-plane FCPM image showing particles embedded in a Lehmann cluster at different depths. (B–D) Vertical cross-sections showing particles localized inside the Lehmann cluster and centered at different $\lambda^{+1/2}$ and $\lambda^{-1/2}$ disclinations. (E) Distance vs. time plot for the motion of a particle perpendicular to the length of the cluster, because it is attracted toward the $\lambda^{+1/2}$ disclination (blue curve; the particle initial position is within the same cholesteric layer as the $\lambda^{+1/2}$ defect line). The red curve shows similar data for the motion of two attracting particles embedded in the same $\lambda^{-1/2}$ disclination. The insets show initial and final particle positions for both cases. (F) Distance dependence of the interaction energy for particle-defect and particle-particle interactions. (G–I) Self-arrangement of colloids within a Lehmann cluster as seen in vertical cross-sectional images perpendicular to (G and H) and along (I) the length of the defect.

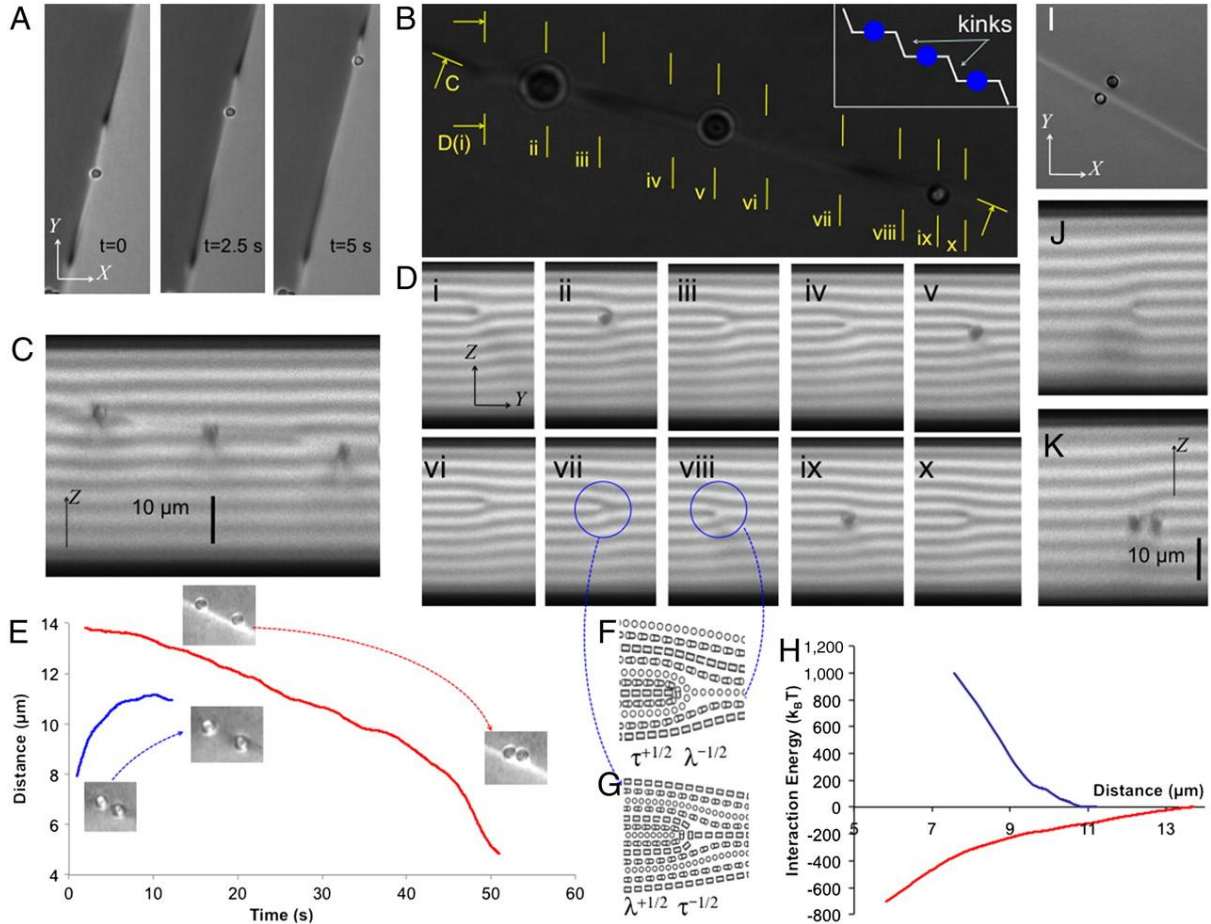

**Fig. 5.** Three-dimensional patterning of particles along edge dislocations with kinks. (A) Lateral manipulation of a kink in an edge dislocation using an optically trapped particle. (B) In-plane image showing particles embedded within an edge dislocation and separated by kinks; the inset schematically shows how colloidal particles localize on edge dislocations at different depths within the layered system while separated by kinks. (C) Vertical FCPM cross-section along the length of an edge dislocation as marked in B. (D) A series of vertical FCPM cross-sections in planes perpendicular to the dislocation as marked in B. (E) Distance vs. time plot describing the motion of interacting particles embedded into the dislocation: The particles repel from each other (blue curve) when separated by a kink but mutually attract (red curve) without kinks. The initial and final positions of the particles are shown in the insets for both cases. (F and G) Schematics of two alternating core structures of the $b=p/2$ dislocation as it shifts vertically to form kinks. (H) The distance dependence of the interaction energy for both the repulsive (blue) and the attractive (red) interparticle interactions within the edge dislocation with and without separating kinks, respectively. (I) Planar and (J and K) vertical cross-sections of particles arranged in a dipolar fashion within the core of a $b=p/2$ dislocation.

# Supporting Information

**Integrated Three-Dimensional Optical Manipulation and Imaging Setup**

The integrated system for simultaneous optical manipulation and 3D imaging is built around an inverted microscope (IX 81 Olympus) (Fig. S1). The holographic optical trapping (HOT) part of the setup is centered on a reflective, electrically addressed, phase-only spatial light modulator (SLM) obtained from Boulder Nonlinear Systems (XY series, P512-1064). The SLM has 512×512 pixels, each 15×15 $\mu m^2$ in size. The HOT setup employs an Ytterbium-doped fiber laser (YLR-10-1064, IPG Photonics) operating at 1,064 nm. The laser is linearly polarized with a Glan-laser polarizer and the polarization direction is adjusted with a half-wave retardation plate to optimize the phase modulation efficiency of the SLM. Before the beam is incident on the SLM, it is expanded to overfill the active area of the SLM and then is resized so as to overfill back aperture of the objective. The SLM controls the phase of the beam on a pixel-by-pixel basis according to the computer-generated holographic patterns supplied at a refresh rate of 30 Hz for the entire pixel array. This spatially phase-modulated beam is imaged at the back aperture of the microscope objective, which recreates the three-dimensional (3D) spatial trap pattern in the sample. A dichroic mirror DM-IR (Chroma Technology Corp.) reflects the trapping beam at 1,064 nm while allowing visible light (used for imaging purposes) to transmit through it to the confocal microscopy scanning head and charge-coupled device (CCD) camera. For FCPM imaging, we use laser-scanning fluorescence confocal unit (FV300, Olympus), which scans the excitation laser beam at 488 nm with galvano-mirrors. For 2PEF-PM imaging, we have employed a tunable (680–1,080 nm) Ti:Sapphire oscillator (Chameleon Ultra II, Coherent) emitting 140-fs pulses at the repetition rate of 80 MHz. We tune the wavelength to 980 nm for the two-photon excitation of the dye BTBP. The 2PEF-PM signal from the dye is collected in epi-detection mode with a photomultiplier tube (H5784-20, Hamamatsu) and a series of interference filters. We use oil-immersion objectives with high numerical aperture (NA), 60× (NA = 1.42) and 100× (NA = 1.4), both obtained from Olympus. The same objectives are used for imaging as well as optical trapping. The integrated setup for holographic optical manipulation and 3D imaging is described in more detail elsewhere (1, 2).

**Computer Simulations of Director Structures Within and Around Defects**

We use numerical minimization of Frank elastic free energy to obtain the static equilibrium structure of oily streaks with different number of interrupted layers and defect cores composed of nonsingular λ-disclinations. The free energy for a cholesteric LC of pitch $p$ is given by

$$F_{elastic} = \int \left\{ \frac{K_{11}}{2}(\nabla \cdot \hat{n})^2 + \frac{K_{22}}{2}\left[\hat{n} \cdot (\nabla \times \hat{n}) + \frac{2\pi}{p}\right]^2 + \frac{K_{33}}{2}[\hat{n} \times (\nabla \times \hat{n})]^2 \right\} dV, \quad (S1)$$

where $K_{11}$, $K_{22}$, $K_{33}$ are elastic constants describing the elastic energy cost of splay, twist, and bend deformations of $\mathbf{n}(\mathbf{r})$, respectively. The minimization of the free energy to find the equilibrium director field is performed using the director relaxation method (1, 3–4). The experimentally reconstructed director orientation pattern is used to set the initial and boundary conditions. To find the equilibrium structure in the vertical cross-sectional plane ($xz$ plane, perpendicular to layers and the defect lines), the functional derivatives of the total free-energy $F_{elastic}$, $\delta F_{elastic}/\delta n_{x,z}$ are set to zero. The minimization of free energy has been done for material parameters of nematic hosts 5CB and ZLI-2806. The calculations of line tension of oily streaks using Eq. S1 disregard the $K_{24}$-term (1). Including this term would likely result in slight lowering of the line tension values; however, the values of the $K_{24}$ elastic constant are unknown for the used materials (1). The obtained director structures and line tension values are consistent with experimental results. Using the obtained minimum-energy director structure, we simulate FCPM textures by using the fact that the intensity of the FCPM signal $I_{FCPM} \propto \cos^4\theta$, where $\theta$ is the angle between the director and the polarization of excitation light; this yields computer-simulated textures of $\mathbf{n}(\mathbf{r})$, which closely resemble experimental results (Fig. S4).

**Analytical Estimates**

The elastic free energy density for a lamellar medium can also be given in terms of the principal radii of curvature ($R_1$ and $R_2$) of the lamellar deformation and the dilation/compression of the layers ($\gamma = (d - d_0)/d_0$, the relative difference between the actual and the equilibrium layer thickness, $d$ and $d_0 = p/2$, respectively) as

$$F_{elastic} = \frac{K_1}{2}\left(\frac{1}{R_1} + \frac{1}{R_2}\right)^2 + \frac{1}{2}B\gamma^2. \quad (S2)$$

The constant $K_1$ (describing the splay of the $\chi$-director field) and the Young modulus $B$ (describing the energy cost associated with dilation/compression of layers) are related to the Frank elastic constants of the liquid crystal as $K_1 = 3K_{33}/8$ and $B = K_{22}(2\pi/p)^2$. Assuming that the layer spacing within and around the oily streak and dislocation defects remains intact, the line tension of these defects can be calculated as the lamellar curvature energy per unit length (5, 6)

$$T = \sum_{i=1}^{m} \frac{\pi K_1(p/2)}{r_i} + \sum_{i=m+1}^{\infty} \frac{K_1 p}{r_i} \sin^{-1} \frac{s/2}{r_i}, \tag{S3}$$

where $r_i = (p/2)[1/2 + (i-1)]$ and $s$ is the width of an oily streak defined as the distance between constituent $\lambda^{+1/2}$-disclinations. The analytical expression (S3) is based on the assumption of equidistant layers and can be used only for rough estimates of tension. A comparison of computer simulations and analytical estimates with Eq. S3 indicates that the core energy of oily streaks cannot be disregarded (Table S2), even in the case when the core is composed of only nonsingular disclinations.

**Experimental Estimates of Defect Line Tension**

The defect tension is experimentally measured using several different approaches, with the use of colloidal particles as "handles" and without them. In the first method, using optical tweezers, a particle embedded within the defect is pulled orthogonally to the defect line, until a resultant defect is engendered. Upon releasing the particle from the trap, this newly created defect contracts, dragging the particle along with it (Movies S1–S4). The motion of the particle is recorded at the rate of 15 frames per second using a charge-coupled device camera (Flea, PtGrey, 640×480 pixels) and analyzed with particle tracking software (ImageJ or Tracker) to yield particle velocities and hence the tensile forces acting on the particles. An alternative approach involves manipulation of a defect line by pushing an optically trapped particle orthogonally to the length of the defect. At a fixed value of optical power, component of the tensile force of the defect will balance the trapping force up to a maximum bending angle, from which the value of tension can be obtained as, $T = F_{trap}/(2\cos\theta)$ (Figs. S5 and S6 and Movie S7). We measured the tension of defects of different widths as shown in Fig. S6. Because the wider defects are also more vertically spread, while the particle size is the same (4 µm), the value of tension obtained thus for defects with large s is typically an underestimation. To directly probe the colloidal pair interactions or defect-colloid interactions described in this work (7), the particles are first positioned at appropriate lateral distances and axial location in their initial configuration using optical tweezers. The laser traps are then switched off and we track the particle motion under the effect of interacting forces balanced by the viscous drag force. Because both particles are free to move, the interaction force on each particle is then equal to the defect line tension and calculated as $F_{int} = (1/2)F_{viscous} = 3\pi\eta R v$.

In addition to methods described above, that involve the use of optically trapped particles as handles, we also obtain the line tension values of defects by directly optically manipulating them with the laser tweezers. For this, we use a beam having power high enough for the electric field of the focused laser beam to reorient $\mathbf{n}(\mathbf{r})$ within the focal volume. This region can then be trapped, because for the particular polarization of trapping beam, it has effective refractive index ($\approx n_e$) higher than that of the surrounding LC and because of the involved attractive elastic forces between the defect and laser-induced distortion (8). This allows direct manipulation of the defect without using a particle. Upon turning off the trap, the extended defect structure shrinks (Figs. S2 and S3 and Movie S5) and the tip of the defect is tracked to measure its rate of contraction. The rate of contraction in this case is much higher compared to the case when a particle is attached at the tip, because there is no viscous drag acting on a colloid in this case. This transformation of the defect involves a continuous transformation of the director structure at the defect tip. The rate of contraction of the defect depends on its width $s$, the rotational viscosity coefficient ($\gamma_1$), and the twist elastic constants ($K_{22}$) of the LC, scaling as $4K_{22}/s\gamma_1$ (6). For example, using this rough analytical estimation, one obtains a value of the contraction rate of 37.3 µm/s for a defect with $s = 1.5p$ in 5CB, which is consistent with the experimental value of 39.9 µm/s (Fig. S2). The value of tension obtained from the measured rate of contraction can be estimated as approximately $(\pi/4)s\gamma_1 v$ and is about 23.6 pN for a defect with $s = 1.5p$ in 5CB-based cholesteric LC, consistent with the values measured using the two different colloidal-particle-based methods of 23.1±0.7 pN (Fig. S6A) and 20.5 pN (Fig. 2A).

**Supplementary References**

**Supplementary Figures**

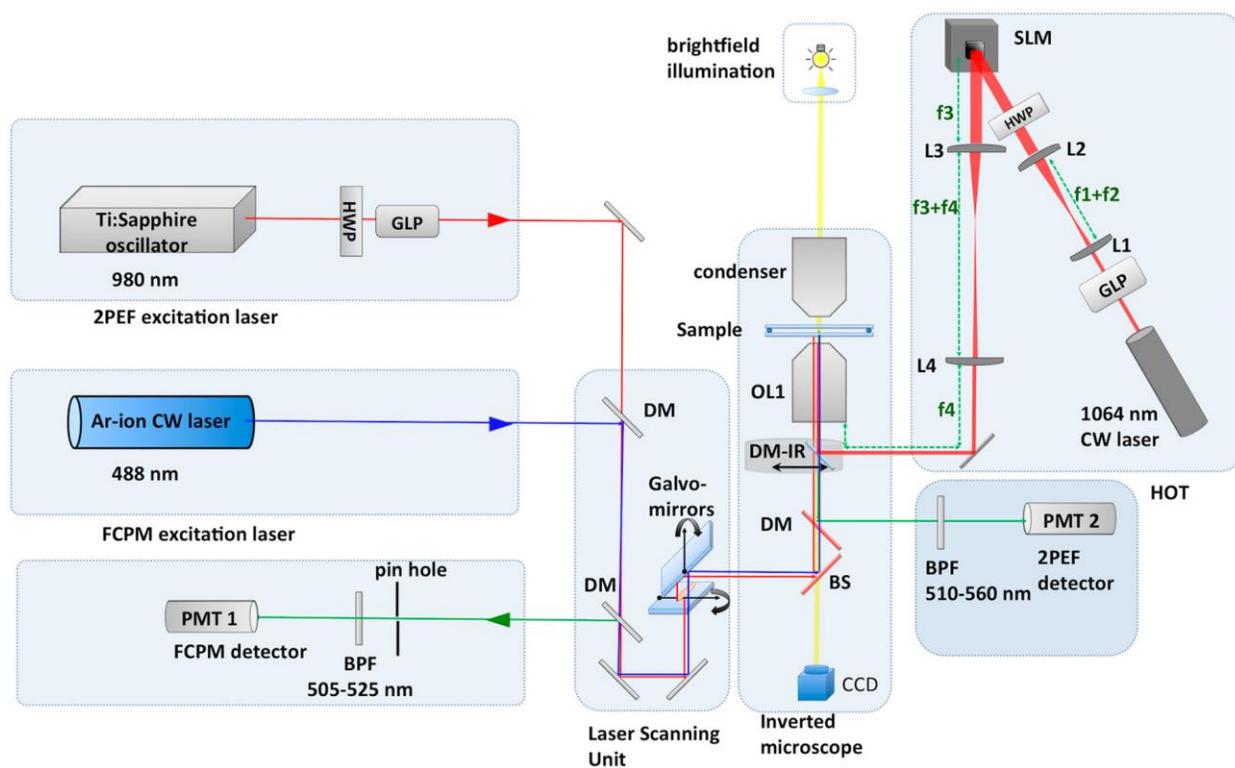

**Fig. S1.** A schematic of the integrated 3D optical manipulation and imaging setup. HWP: half-waveplate, GLP: Glan laser polarizer, DM: dichroic mirror, BPF: bandpass filter, OL1: objective lens, SLM: spatial light modulator, L1–L4: plano-convex lenses. f1–f4: focal length of the lenses, PMT: photomultiplier tube, BS: beam splitter.

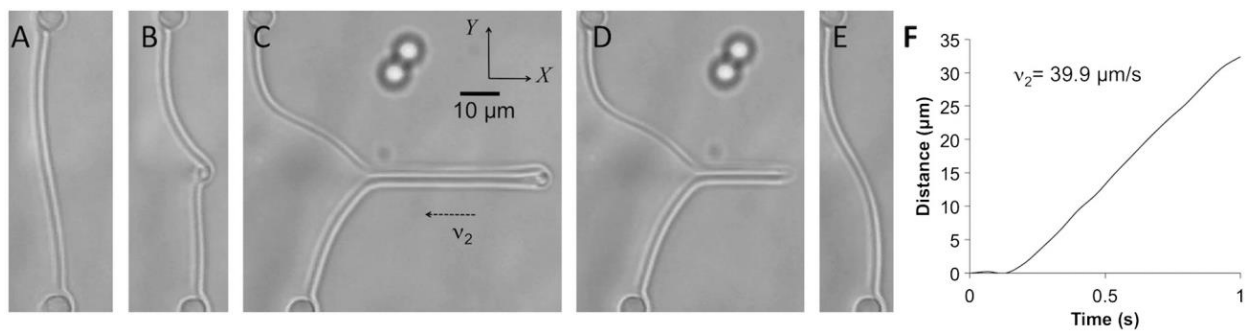

**Fig. S2.** Measurement of defect line tension by means of its direct optical manipulation. (A–C) A defect structure is drawn out of a Lehmann cluster with the help of optical tweezers. (C–E) When the trap is released, the defect structure contracts to reduce its tensile free energy. (F) Videomicroscopy tracking of the tip of the contracting defect.

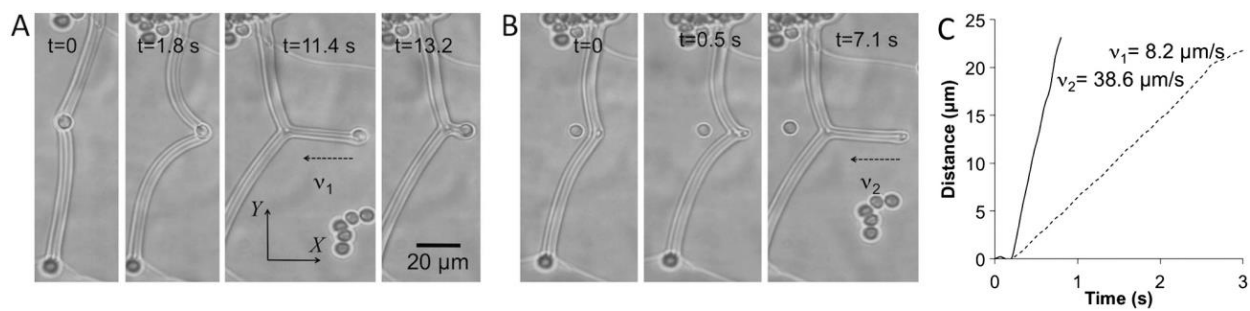

**Fig. S3.** Measurement of line tension of a defect with multiple interrupted layers (large s) by means of its manipulation using a particle as a "handle." (A) A colloidal particle embedded inside an oily streak is manipulated to "draw" a new defect. Upon turning off the trap, the particle returns to its original position under the tension of the stretched defect line. (B) Similar manipulation without the help of a particle, performed directly on the defect. (C) Contraction of the defect vs. time in each case.

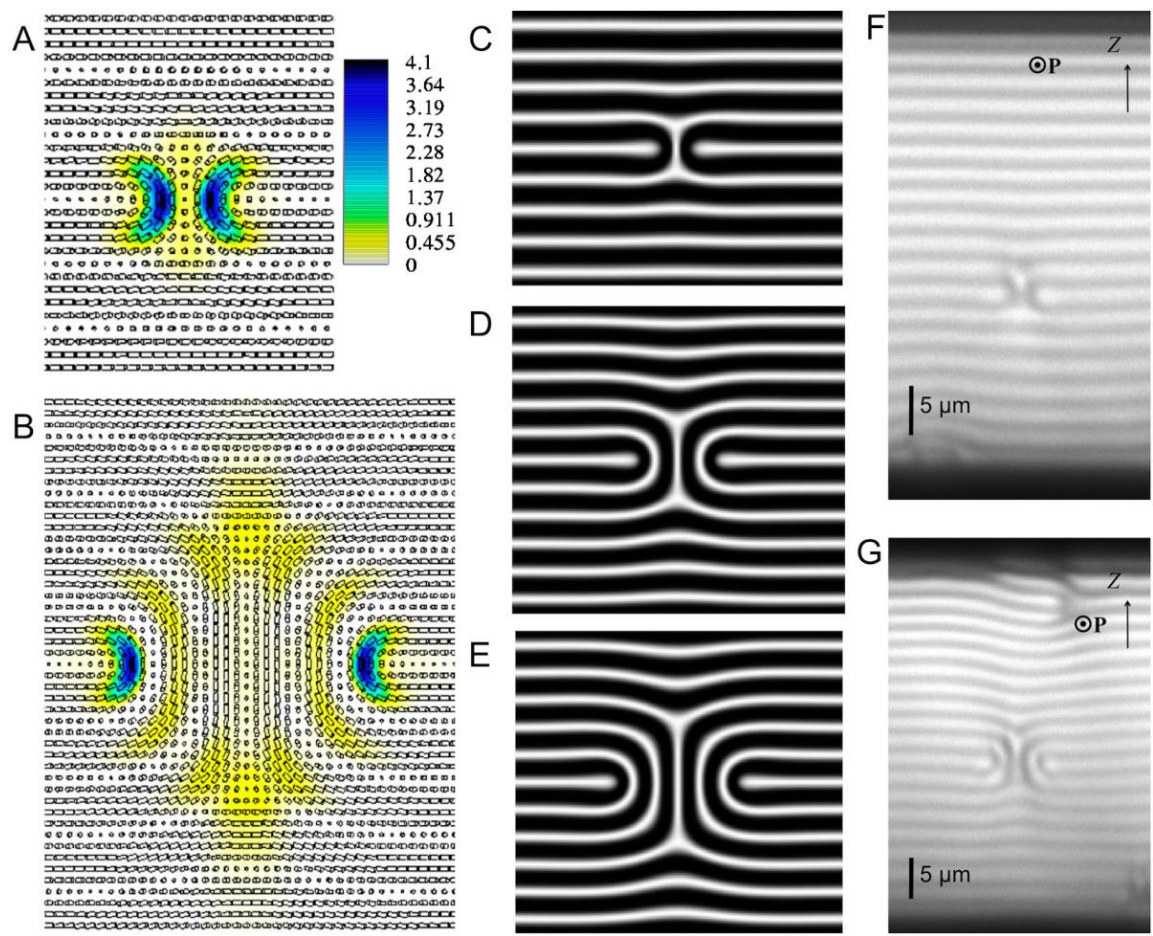

**Fig. S4.** Computer simulations and 3D imaging of the director structure of oily streaks. (A and B) Director configurations and the corresponding relative Frank elastic free-energy density for oily streaks of width $s \approx p$ and $s \approx 3p$ in the vertical cross-section; the spatial pattern of the Frank elastic free energy density is calculated using Eq. S1 and shown by means of the color-coded scale with energy density ranging from that of the minimum-energy uniform cholesteric structure (white) to that of highly distorted regions with highest energy cost (black). (C–E) Computer-simulated FCPM cross-sectional images for oily streaks with $s \approx p$, $\approx 2p$, and $\approx 3p$, respectively. (F and G) Experimental FCPM cross-section obtained for oily streaks of width $s \approx p$ and $\approx 2p$, respectively.

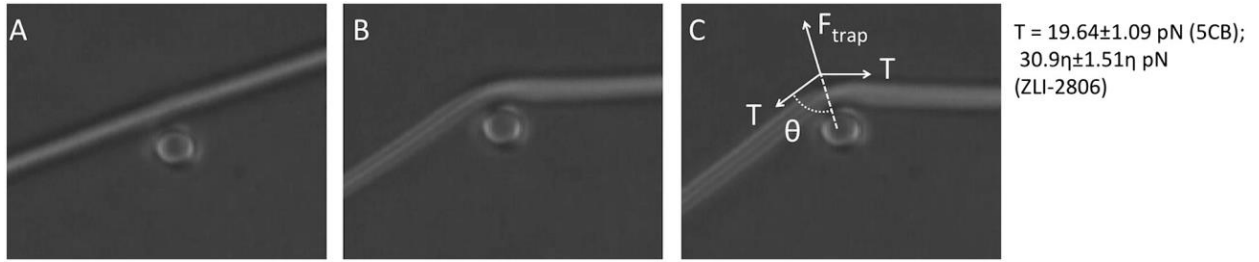

**Fig. S5.** Measurement of tension of a Lehmann cluster by its manipulation using an optically trapped colloid. (A) The defect and the particle in the initial position. (B) The optically trapped particle is pushed against the defect until (C) the component of the line tension balances the laser trapping force, yielding the line tension of about $T = 19.64$ pN, determined using the maximum bending angle $\theta$. The measurement was done for 5CB- and ZLI-2806-based cholesteric LCs with 5 μm pitch, where $\eta$ is (unknown) viscosity of ZLI-2806 in Pa s.

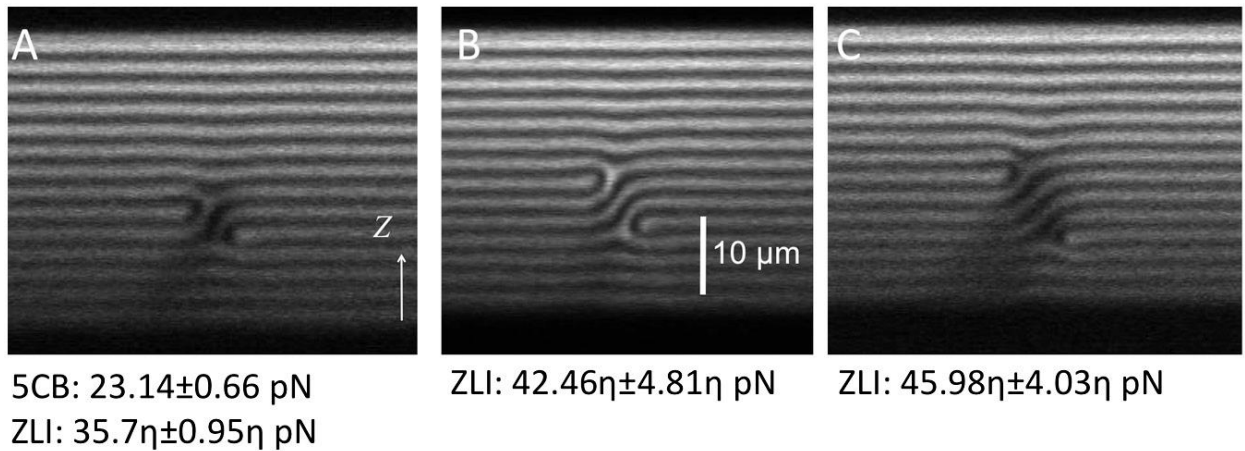

5CB: 23.14±0.66 pN
ZLI: 35.7η±0.95η pN

ZLI: 42.46η±4.81η pN

ZLI: 45.98η±4.03η pN

**Fig. S6.** Tension of defects of different width for cholesteric liquid crystals based on 5CB and ZLI-2806. The cores of the defect are separated by $s = 1.5p$ (A), $2p$ (B), and $2.5p$ (C), respectively. The values of tension are given in terms of the viscosity $\eta$ of ZLI-2806 in Pa s.

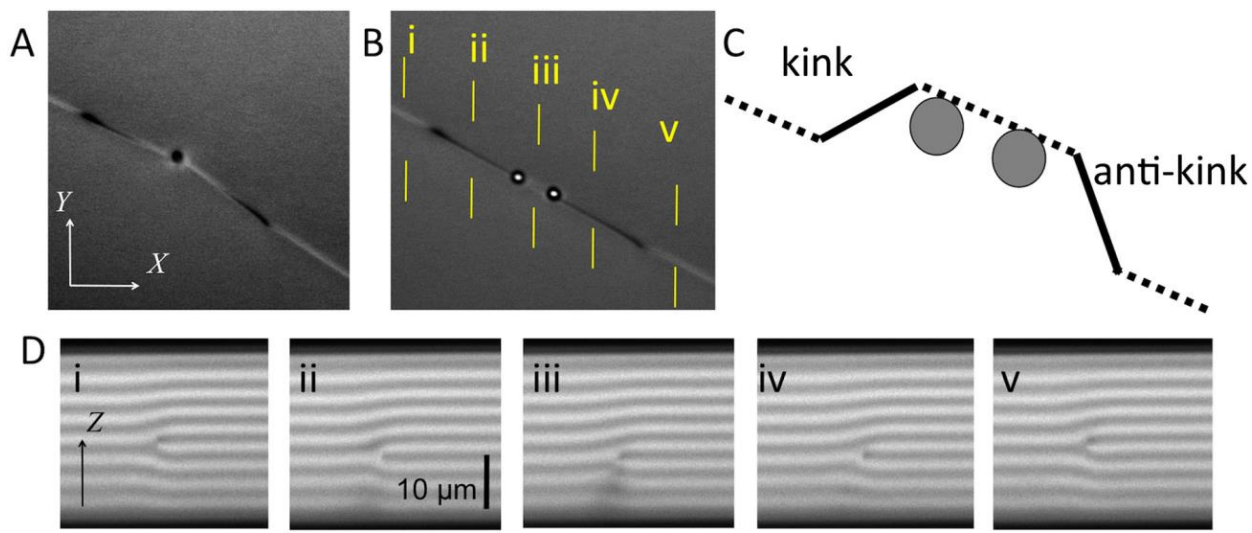

**Fig. S7.** Generation of kink/anti-kink pairs using optical tweezers. (A) A particle embedded inside the $\tau^{+1/2}$ disclination of the dislocation is optically trapped. At high power, the axial laser trapping force acting on the particle pushes the dislocation to a different cholesteric layer. (B) Similar generation of the kink/anti-kink pair can be achieved with axial manipulation of two colloidal particles, which now allow one to control locations of the kink and anti-kink. (C) Schematic illustration showing the controlled generation of a kink/anti-kink pair by simultaneous axial translation of colloidal particles across lamellae. (D) Vertical 2PEF-PM cross-sections of the cholesteric layered sample in the vicinity of the generated kink/anti-kink pair, showing how the axial shift of the dislocation is accompanied by the transformation of dislocation cores.

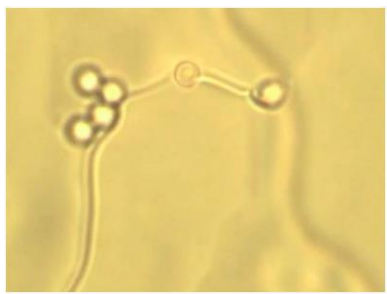

**Movie S1.** Optical generation of defect–particle configurations. The initial structure consists of three particles connected by two separate Lehmann clusters (the center one being free and the outer particles fixed). By optically moving the free particle using laser tweezers, we create a configuration in which the particles are connected by means of a defect node. Movie S1 (MOV)

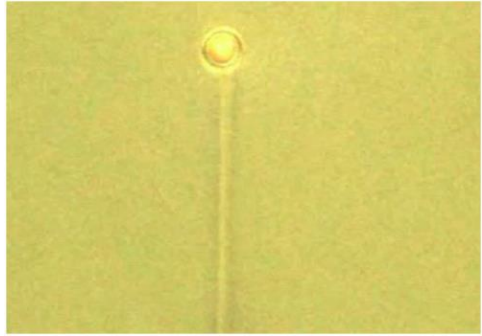

**Movie S2.** A colloidal particle pinned at the tip of a Lehmann cluster is dragged under the effect of the defect line tension as the defect contracts. Movie S2 (MOV)

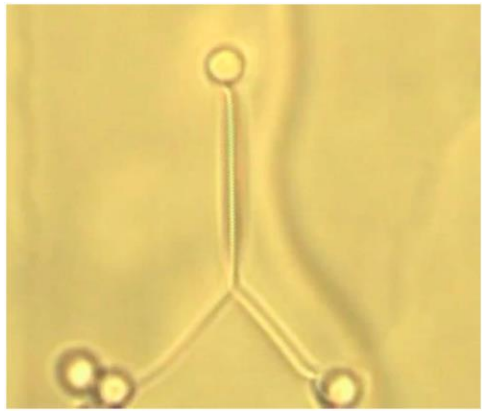

**Movie S3.** A defect node formed by three Lehmann clusters is supported between two stationary and one mobile particles. One of the particles is optically manipulated and then is dragged under the defect tension when released from the laser trap (sequel to Movie S1). Movie S3 (MOV)

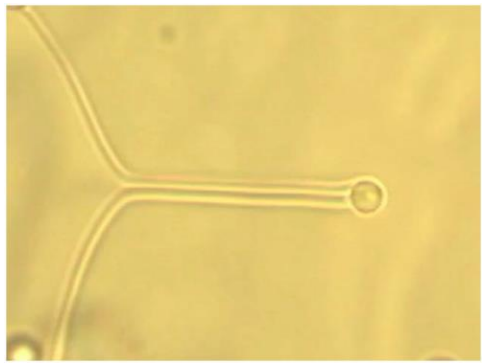

**Movie S4.** A particle embedded inside a Lehmann cluster is optically manipulated to generate a new defect structure. When the particle is released from the optical trap, it is dragged under the tension of the retracting defect. Movie S4 (MOV)

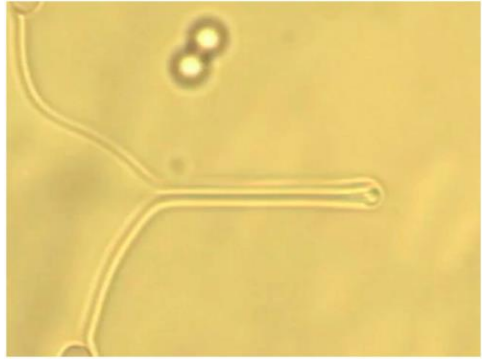

**Movie S5.** Direct optical manipulation of an oily streak generates a new defect structure. This optically generated defect then contracts once released from the laser trap. Movie S5 (MOV)

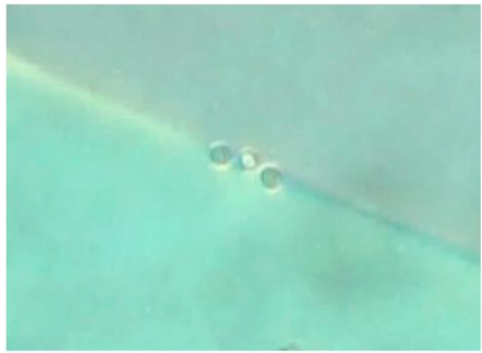

**Movie S6.** Kink-mediated repulsion between colloidal particles embedded into an edge dislocation. The particles are brought close to each other by optically trapping them. Once the traps are turned off, the particles repel from each other. Movie S6 (MOV)

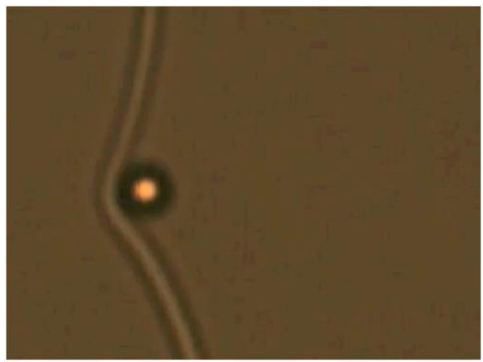

**Movie S7.** An optically trapped particle is used to manipulate a defect line. The maximum value of the bending angle at a given optical power yields the tension of the defect line. Movie S7 (MOV)

**Table S1. Material parameters of used nematic hosts and chiral additives**

| LC | $K_{11}$ pN | $K_{22}$ pN | $K_{33}$ pN | K pN | $n_o$ | $n_e$ | $H_{HTP}$ of CB15 μm$^{-1}$ | $\gamma_1$ mPa·s |
|---|---|---|---|---|---|---|---|---|
| 5CB | 6.4 | 3 | 10 | ≈7 | 1.536 | 1.714 | +7.3* | ≈100 |
| ZLI-2806 | 14.9 | 7.9 | 15.4 | ≈11 | 1.475 | 1.518 | +5.9* | ≈240 |

*The positive sign indicates right-handed sense of twist.

**Table S2. Line tension of oily streaks with different width for cholesteric LC with pitch $p = 5\ \mu m$**

| s | 5CB | | ZLI-2806 | |
|---|---|---|---|---|
| | Simulated (pN) | Eq. S3 (pN) | Simulated (pN) | Eq. S3 estimate (pN) |
| p | 20.5 | 30.9 | 35.1 | 47.6 |
| 2 p | 34.1 | 39.3 | 60.4 | 60.5 |
| 3 p | 47.1 | 44.1 | 86.3 | 68.0 |